\documentclass{PoS}
\usepackage{enumerate}

\title{Light charged Higgs boson production at future $ep$ colliders}

\ShortTitle{Light charged Higgs boson production at future $ep$ colliders}

\author{O. Flores-S\'anchez\\
      Departamento de Sistemas y Computaci\'on
Instituto Tecnol\'ogico de Puebla,
Av. Tecnol\'ogico num. 420 Col. Maravillas,	Puebla, Puebla, C.P. 72220, M\'exico.\\
        E-mail: \email{omar.flores@itpuebla.edu.mx}}

\author{\speaker{J.  Hern\'andez-S\'anchez }\thanks{Supported by SNI-CONACYT, VIEP-BUAP and PRODEP, M\'exico.}\\
        Author affiliation\\
        E-mail: \email{jaime.hernandez@correo.buap.mx}}

\author{C. G. Honorato\\Facultad de Ciencias de la Electr\'onica, Benem\'erita Universidad Aunt\'onoma de Puebla, Apdo. Postal 542, C.P. 72570 Puebla, Puebla, M\'exico.\\
         E-mail: \email{carlosg.honorato@correo.buap.mx}}

\author{S. Moretti\thanks{Supported through the NExT Institute, STFC CG grant ST/L000296/1 and  H2020-MSCA-RISE-2014 grant no. 645722
(NonMinimalHiggs).} \\
School of Physics and Astronomy, University of Southampton, Highfield, Southampton SO17 1BJ, United Kingdom and Particle Physics Department, Rutherford Appleton Laboratory, Chilton, Didcot, Oxon OX11 0QX, United Kingdom \\
E-mail: \email{s.moretti@soton.ac.uk}} 
       
\author{S. Rosado\\
        Facultad de Ciencias F\'{\i}sico-Matem\'aticas,Benem\'erita Universidad Aut\'onoma de Puebla, Apdo. Postal 1364, C.P. 72570 Puebla, Puebla, M\'exico.\\
        E-mail: \email{sebastian.rosado@gmail.com}}

\abstract{We present a recent study of  light charged Higgs boson ($H^-$) production at  the Large Hadron electron Collider (LHeC). We study  the charged current production process $e^- p \to \nu_e q H^-$, taking in account the decay channels  $H^- \to b\bar{c}$ and $H^-\to \tau \bar{\nu}_\tau$. We analyse the process in the framework of the 2-Higgs Doublet Model Type-III (2HDM-III), assuming a four-zero texture in the Yukawa matrices and a general Higgs potential. We consider a variety of both reducible and irreducible backgrounds for the signals of the $H^-$ state. We show that the detection of a light charged Higgs boson is feasible, assuming for the LHeC  standard energy and luminosity conditions.}

\FullConference{XXVII International Workshop on Deep-Inelastic Scattering and Related Subjects - DIS2019\\
		8-12 April, 2019\\
		Torino, Italy}

\begin{document}

\section{Introduction}
After of the discovery of a neutral Higgs boson by the CMS \cite{cms} and ATLAS \cite{Atlas} experiments, practically, the
Standard Model  (SM) has been fully established. However,  in several  extensions of the Higgs sector Beyond the SM (BSM) which can reproduce the SM-like limit of  Electro-Weak Symmetry Breaking (EWSB) using doublet Higgs fields, there appears at least one charged Higgs boson, like in the  2HDM \cite{Branco:2011iw}. Amongst the possible $H^\pm$ decay channels.  the importance of the  $H^\pm \to  cb $ one has been pointed out as a possible viable signal in some models and its detection possibilities have been analysed for the LHC already several years ago  \cite{DiazCruz:2009ek,HernandezSanchez:2012eg,Akeroyd:2016ymd,Akeroyd:2012yg} and very recently for the LHeC \cite{Flores-Sanchez:2018dsr,Hernandez-Sanchez:2016vys} as well. Our own studies have been carried out  in the context of the 2HDM-III, where both Higgs doublets are coupled to both up- and down-type quarks  and  Flavour Changing Neutral Currents (FCNCs) can be controlled by a particular texture in the Yukawa matrices \cite{DiazCruz:2009ek,HernandezSanchez:2012eg,Felix-Beltran:2013tra}. In this work, we present a new analysis of the signals $H^- \to b \bar{c}$ and $H^-\to \tau \bar{\nu}_\tau$ from the process 
 $e^- p \to \nu_e q H^-$ at the LHeC machine, considering the most recent constraints from experimental data \cite{Flores-Sanchez:2018dsr}. In the process $e^- p \to \nu_e q H^-$, $q$ can be a light quark $q_l = u, \, d, \, c, \, s$ or a $b$-quark, with the production stage followed by $H^- \to b\bar{c}$ and $H^-\to \tau \bar{\nu}_\tau$ (Fig. 1), assuming a leptonic decay of the $\tau$
 into an electron or muon. When the final state  is  $H^- \to b \bar{c}$, the main backgrounds are $\nu 3 j$, $\nu 2b j$, $\nu 2j b$ and $\nu t b$ (Fig. 2). For the final state $H^-\to \tau \bar{\nu}_\tau$, these are $\nu j \ell \nu$ and $\nu b \ell \nu$ (Fig. 3).  
 \begin{figure}
        \centering
        \includegraphics[scale = .13]{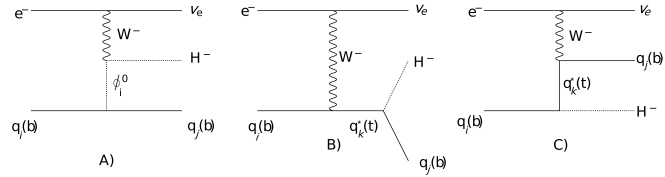}
        \caption{
		Feynman diagrams for the $e^-\ p\rightarrow \nu_e H^- q$ process.  Here, $\phi^0_i=h,H,A$, i.e., any of the neutral Higgs bosons of the BSM scenario considered here. }
        \label{Feynsignal}
\end{figure}

\begin{figure}
        \centering
        \includegraphics[scale = .13]{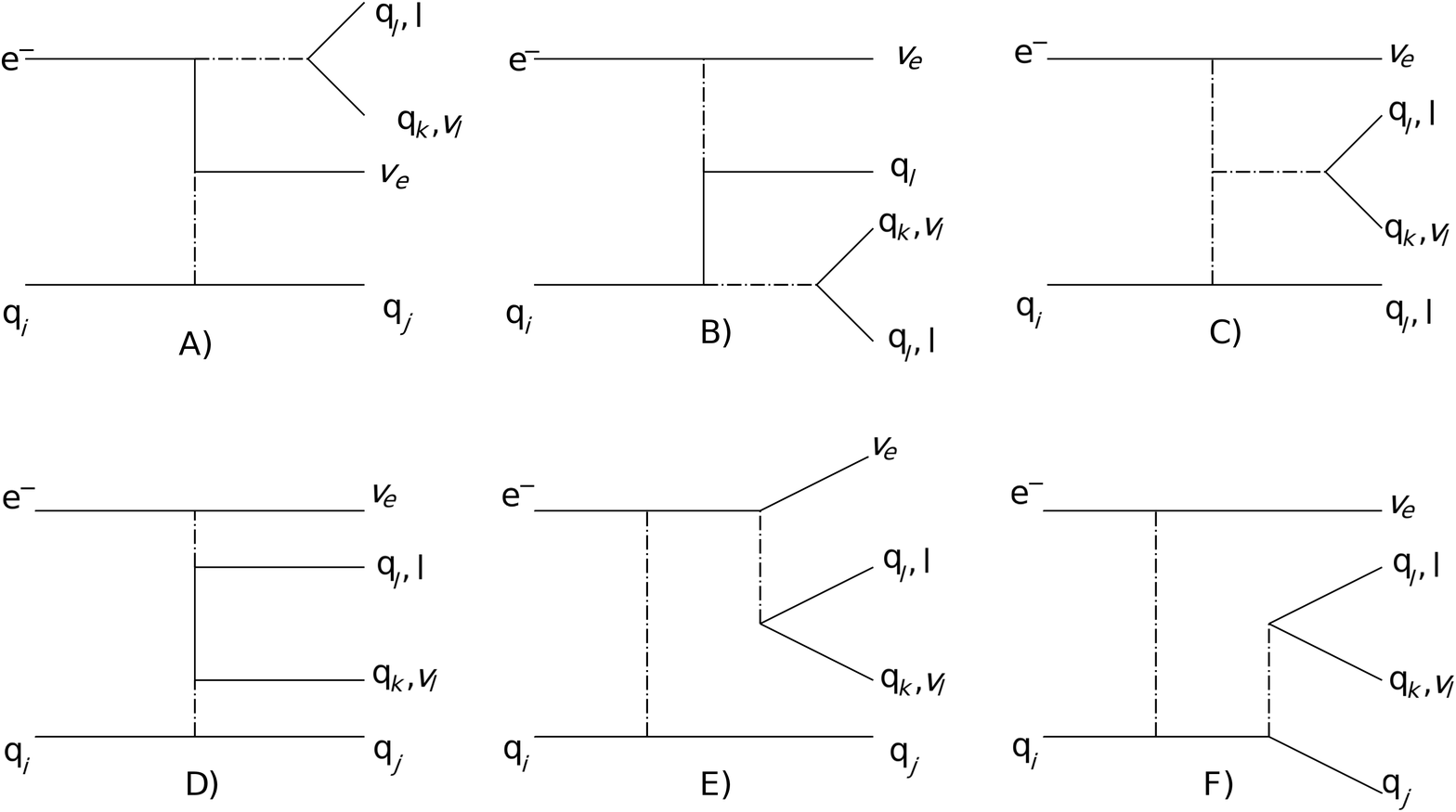}
        \caption{
		Feynman diagrams for the $\nu_e j j j$, $\nu_e b j j$ and $\nu_e b b j$ backgrounds (the change $ q_l \leftrightarrow l $ and $ q_k \leftrightarrow \nu_l $ represents the $\nu_e \nu_l l j$ and $\nu_e \nu_l l b$ backgrounds). Dash-dot lines represent boson fields:  (pseudo)scalars and EW gauge bosons.  }
        \label{FeynBG1}
\end{figure}

\begin{figure} 
        \centering
        \includegraphics[scale = .13]{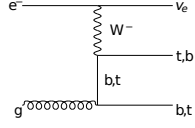}
        \caption{
		Feynman diagrams for the $\nu_e b t$ background. }
        \label{FeynBG2}
\end{figure}
 The plan of this paper is: we present the 2HDM-III in the next section,   then show our results and finally conclude.

 \section{2HDM-III}

 For the 2HDM-III,  a four-zero-texture is implemented and FCNCs are controlled.  Then  the most general $SU(2)_L \times U(1)_Y$ invariant scalar potential for two scalar doublets, $\Phi_i^\dagger= (\phi_i^-, \phi_i^{0*})$ ($i=1, \, 2$),   is considered, which is
\begin{eqnarray}
V(\Phi_1,\Phi_2)&=&\mu_1^2(\Phi_1^\dagger \Phi_1)+\mu_2^2(\Phi_2^\dagger \Phi_2)-\Big(\mu_{12}^2(\Phi_1^\dagger \Phi_2)+h.c.\Big)+\frac{1}{2}\lambda_1(\Phi_1^\dagger \Phi_1)^2+\frac{1}{2}\lambda_2(\Phi_2^\dagger \Phi_2)^2\nonumber\\
&&+\lambda_3(\Phi_1^\dagger \Phi_1)(\Phi_2^\dagger \Phi_2)+\lambda_4(\Phi_1^\dagger \Phi_2)(\Phi_2^\dagger \Phi_1)+\left[\frac{1}{2}\lambda_5(\Phi_1^\dagger \Phi_2)^2+\Big(\lambda_6(\Phi_1^\dagger \Phi_1)\right.\nonumber\\
&&\left.+\lambda_7(\Phi_2^\dagger \Phi_2)\Big)(\Phi_1^\dagger \Phi_2)+h.c.\right], \label{potencial}
\end{eqnarray}
where we assume that all parameter of Higgs potential are real, including the Vacuum Expectation Values (VEVs) of the Higgs  fields, $v_{1,2}$. The Yukawa Lagrangian is:
\begin{equation}\label{yukawa-lagran}
{\cal L}_Y=-\Big(Y_1^u\bar Q_L \tilde \Phi_1 u_R+Y_2^u \bar Q_L\tilde \Phi_2 u_R+Y_1^d\bar Q_L \Phi_1 d_R+Y_2^d\bar Q_L \Phi_2 d_R+Y_1^l\bar L \Phi_1 l_R+Y_2^l \bar L_L \Phi_2 l_R\Big),
\end{equation}
where $\tilde \Phi_{1,2}= i \sigma_2 \Phi_{1,2}^*$. The fermion mass matrices after  
EWSB are expressed by: 
$M_f = \frac{1}{2} (v_1 Y_1^f + v_2 Y_2^f )$, $f=u, \, d, \, l$, assuming that both Yukawa matrices  $Y_1$ and $Y_2$ have the four zero-texture form  and are Hermitian  \cite{DiazCruz:2009ek,HernandezSanchez:2012eg,Felix-Beltran:2013tra}. Upon diagonalising the mass matrices, one obtains the rotated matrix $Y_n^f $: 
$[\tilde Y_n^f]_{ij}=\frac{\sqrt{m_i^fm_j^f}}{v}[\tilde\chi_n^f]_{ij}=\frac{\sqrt{m_i^fm_j^f}} {v}[\chi_n^f]_{ij}e^{i\theta_{ij}^f}$,
where the $\chi$  parameters  can be constrained by flavour physics \cite{HernandezSanchez:2012eg,Felix-Beltran:2013tra}, with $v= \sqrt{v_1^2+v_2^2}$. In agreement with Ref. \cite{HernandezSanchez:2012eg}, one can get a generic expression for the fermionic couplings of the charged Higgs bosons:
\begin{eqnarray}
{\cal L}^{\bar f_i f_j \phi}&=&-\left\{\frac{\sqrt{2}}{v}\bar u_i(m_{d_j}X_{ij}P_R+m_{u_i}Y_{ij}P_l)d_jH^++\frac{\sqrt{2}m_{l_j}}{v}Z_{ij}\bar{v_L}l_RH^++h.c.\right\}, \label{Yukawa-Charged}
\end{eqnarray}
 where 
$X_{ij}$, $Y_{ij}$ and $Z_{ij}$ are defined as follows:
\begin{eqnarray}
X_{ij} &=& \sum_{l=1}^{3} \left( V_{\rm CKM} \right)_{il} \left[ X \frac{m_{d_l}}{m_{d_j}} \delta_{lj} -\frac{f(X)}{\sqrt{2}} \sqrt{\frac{m_{d_l}}{m_{d_j}}} \tilde{\chi}_{lj}^{d} \right], \\
Y_{ij} &=& \sum_{l=1}^{3} \left[ Y \delta_{il} -\frac{f(Y)}{\sqrt{2}} \sqrt{\frac{m_{u_l}}{m_{u_i}}} \tilde{\chi}_{il}^{u} \right] \left( V_{\rm CKM} \right)_{lj}, \\
Z_{ij}^{l} &=& \left[ Z \frac{m_{l_i}}{m_{l_j}} \delta_{ij} -\frac{f(Z)}{\sqrt{2}} \sqrt{\frac{m_{l_i}}{m_{l_j}}} \tilde{\chi}_{ij}^{l} \right],
\end{eqnarray}
where 
$f(x)=\sqrt{1+x^2}$  and the parameters $X$, $Y$ and $Z$   are arbitrary complex numbers, which can be related to $\tan \beta$ or $\cot \beta$ when $\chi_{ij}^f=0$ \cite{HernandezSanchez:2012eg}, thus  one can recovers the standard four types  of the 2HDM (Tab.~\ref{XYZ})\footnote{We will call these 2HDM-III `incarnations'  2HDM-III like-$\chi$ scenarios, where $\chi=$ I, II, X and Y.} and one can write  the Higgs-fermion-fermion $(\phi ff)$ couplings  as $g_{\rm 2HDM-III}^{\phi ff} = g_{\rm 2HDM-any}^{\phi ff} + \Delta g$, where $g_{\rm 2HDM-any}^{\phi ff}$ is the coupling $\phi f f$ in any of the 2HDMs with discrete symmetry and $\Delta g$ is the contribution of the four-zero-texture. 
\begin{table}[htp]
\centering
\resizebox{5cm}{!} {
\begin{tabular}{|c|c|c|c|}
\hline
2HDM-III & $X$ & $Y$& $Z$\\
\hline
2HDM Type I & $-\cot \beta$ &  $ \cot \beta$ & $- \cot \beta$ \\
\hline
2HDM Type II & $\tan \beta$ &  $ \cot \beta$ & $ \tan \beta$ \\
\hline
2HDM Type X & $-\cot \beta$ &  $ \cot \beta$ & $ \tan \beta$ \\
\hline
2HDM Type Y  & $\tan \beta$ &  $ \cot \beta$ & $ -\cot \beta$ \\
\hline
\end{tabular}
}
\caption{The parameters $X$, $Y$ and $ Z$ of the 2HDM-III defined in the Yukawa interactions when $\chi_{ij}^f=0$ so as to recover the standard four types of 2HDM.}
\label{XYZ}
\end{table}%

We take four Benchmark points (BPs)  where the decay channels $H^- \to b \bar{c}$ and $H^-\to \tau \bar{\nu}_\tau$  can offer the most optimistic chances for detection \cite{Flores-Sanchez:2018dsr}. 
\begin{itemize}
\item Scenario 2HDM-III like-I: $\cos (\beta- \alpha)= 0.5$, $\chi^u_{22}= 1$, $\chi^u_{23}= 0.1$, $\chi^u_{33}= 1.4$, $\chi^d_{22}= 1.8$, $\chi^d_{23}= 0.1$,
$\chi^d_{33}= 1.2$,  $\chi^\ell_{22} =-0.4, \chi^\ell_{23}= 0.1$, $\chi^\ell_{33} =1$ with $Y \gg X, \, Z$.
\item Scenario 2HDM-III like-II: $\cos (\beta- \alpha)= 0.1$, $\chi^u_{22}= 1$, $\chi^u_{23}= -0.53$, $\chi^u_{33}= 1.4$, $\chi^d_{22}= 1.8$, $\chi^d_{23}= 0.2$,
$\chi^d_{33}= 1.3$,  $\chi^\ell_{22} =-0.4, \chi^\ell_{23}= 0.1$, $\chi^\ell_{33} =1$ with $X, \, Z \gg Y$.
\item Scenario 2HDM-III like-X: the same parameters of scenario 2HDM-III like-II but  $Z \gg X, \, Y$.
\item Scenario 2HDM-III like-Y: the same parameters of scenario 2HDM-III like-II but  $X \gg Y, \, Z$.
\end{itemize}

\section{Results}

We assume the LHeC standard Centre-of-Mass (CM) energy of $\sqrt{ s_{ep}} \approx 1.3$ TeV and luminosity of $L=100$ fb$^{-1}$. For the signatures $H^- \to b \bar{c}$ and $H^- \to \tau \nu_\tau$ the inclusive event  rates are substantial, of order up to several thousands in all four cases. Some BPs, maximising the signal rates are given in Tab. \ref{tab:BR}. 
\begin{table}[!t]
\centering
\resizebox{13cm}{!} {
		\begin{tabular}{|c|c|c|c|c|c|c|c|c|c|c|c|}
		\hline
		2HDM-III & \multicolumn{3}{|c|}{Parameters} & \multicolumn{4}{|c|}{$\sigma(ep\to\nu_eH^- q)$ (pb)}
		 & BR$(H^-\to b\bar c)$ &  BR$(H^-\to \tau \bar\nu_\tau)$\\
		\cline{2-8}
		like- & ${X}$ & ${Y}$ & ${Z}$ &{$m_{H^\pm}=110$ GeV}&{$130$ GeV}&
		{$150$ GeV} &{$170$ GeV}&{$m_{H^\pm}=110$ GeV}
		&{$m_{H^\pm}=110$ GeV}\\
		\hline
		\hline
		I & {0.5} & {17.5} & {0.5}&{$2.56\times 10^{-2}$} &{$1.30\times 10^{-2}$}
		&{$3.47\times 10^{-3}$}&{$1.35\times 10^{-4}$}&{$9.57\times 10^{-1}$}
		&{$2.5\times 10^{-4}$}\\
		\hline
		II & {20}& {1.5} & {20}& {$2.18\times 10^{-2}$}& {$1.13\times 10^{-2}$}
		& {$2.95\times 10^{-3}$} & {$5.89\times 10^{-5}$} & {$9.9\times 10^{-1}$}& 
		{$2.22\times 10^{-4}$}\\
		\hline
		X & {0.03}& {1.5} & {$-33.33$}& {$6.49\times 10^{-2}$}&{$3.39\times 10^{-2}$}
		& {$8.83\times 10^{-3}$} & {$2.34\times 10^{-4}$} & {$9.28\times 10^{-2}$}& 
		{$9.04\times 10^{-1}$}\\
		\hline
		Y & {13}& {1.5} & {$-1/13$}& {$6.41\times 10^{-2}$}& {$3.27\times 10^{-2}$}
		& {$8.47\times 10^{-3}$} & {$2.2\times 10^{-4}$} & {$9.91\times 10^{-1}$}& 
		{$6.12\times 10^{-3}$}\\
		\hline
				\end{tabular}
		}
		\caption{The BPs that we studied for the 2HDM-III  in  the incarnations like-I, -II, -X and -Y. We present cross sections and $BRs$ at parton level for some $H^\pm$ mass choices. }
		\label{tab:BR}
\end{table}
The scenarios and signatures that we will study are as follows.

\begin{itemize}
\item BPs from 2HDM-III like-I, -II and -Y, where the most relevant decay process is $H^- \to b\bar{c}$, the final state is $3j+E_T \hspace{-.4 cm} / $\hspace{.2 cm}.

\item BP from 2HDM-III like-X, where the most relevant decays process is $H^- \to \tau\bar\nu_\tau$, the final state is  $j+l+E_T \hspace{-.4 cm} / $\hspace{.2 cm},
where $l=e,\mu$ (from a leptonic $\tau$ decay) and the jet is $b$-tagged.
\end{itemize}
We have used CalcHEP 3.7 \cite{Belyaev:2012qa} as parton level event generator, interfaced to the CTEQ6L1 Parton Distribution Functions (PDFs) \cite{Pumplin:2002vw}, then PYTHIA6  \cite{Sjostrand:2006za} for the parton shower, hadronisation and hadron decays and PGS \cite{PGS} as detector emulator, by using a LHC parameter card suitably modified for the LHeC \cite{AbelleiraFernandez:2012cc,Bruening:2013bga}. We considered a calorimeter coverage $|\eta|<5.0$, with segmentation $\Delta\eta\times\Delta\phi=0.0359\times 0314$. Besides, we used Gaussian energy resolution, with $\frac{\Delta E}{E}=\frac{a}{\sqrt{E}}\oplus b$,
where $a=0.085$ and $b=0.003$ for the Electro-Magnetic (EM) calorimeter resolution and $a=0.32$, $b=0.086$ for the hadronic calorimeter resolution, with $\oplus$
meaning addition in quadrature  \cite{AbelleiraFernandez:2012cc,Bruening:2013bga}. The algorithm to perform jet finding was a ``cone" one with jet radius $\Delta 
R=0.5$. The calorimeter trigger cluster finding a seed(shoulder) threshold was 5 GeV (1 GeV). We took $E_T(j)>10 $ GeV for a jet to be considered so, in addition to
the isolation criterion $\Delta R(j;l)>0.5$. Finally, we have mapped the kinematic behaviour of the final state particles using MadAnalysis5  \cite{Conte:2012fm}.

\subsection{The  process  $e^-q\to \nu_e H^- b$  with {$H^-\to b \bar{c}$} for the 2HDM-III  like-I, -II and -Y}

In this subsection we study the final state with one $b$-tagged jet and one light jet (associated with the secondary decay $H^-\to b \bar{c}$).
\begin{enumerate}[I]
\item First, we select only events with exactly three jets in the final state. Then, we reject all events without a $b$-tagged jet. Then we 
keep events like $3j+E_T \hspace{-.4 cm} / $\hspace{.2 cm} with at least one $b$-tagged jet.
\item The second set of cuts is focused on selecting two jets (one $b$-tagged, labelled as $b_{\rm tag}$, and one not, labelled as $j_{\rm c}$) which are central in the detector. First, we demand that   $P_T(b_{\rm tag})>30(40)$ GeV and $P_T(j_{\rm c})>20(30)$ GeV for $m_{H^\pm}=110,130(150,170)$ GeV (here, $P_T$ is the
transverse momentum). Then, we impose a cut on the pseudorapidity $|\eta(b_{\rm tag},j_{\rm c})|<2.5$ of both these jets and, finally, select events in which
$1.8(2)<\Delta R(j_{\rm c};b_{\rm tag})<3.4(3.4)$  in correspondence of $m_{H^\pm}=110,130(150,170)$ GeV (where $\Delta R$ is the standard cone separation).
\item The next cut is related to the selection of a forward third generic jet (it can be either a light jet or a $b$-tagged one).
Our selection for such a third jet is $|\eta|>0.6 $ (with a transverse momentum above 20 GeV).
\item
We then implement the following selection criterium:  
$m_{H^\pm}-20$ GeV $<M(b_{\rm tag},j_{\rm c})< m_{H^\pm}$. Finally, considering the presence of a hadronic $W^\pm$ boson decay, we impose that
$M (j_{\rm c},j_{\rm f})>80$ GeV or $M(j_{\rm c},j_{\rm f})<60$ GeV (where $j_{\rm f}$ labels the forward jet).
\end{enumerate}

\begin{table}
\centering
\resizebox{10cm}{!} {
	\label{cuts1}
		\begin{tabular}{|r|c|c|r|r|r|r|c|}
		\hline
		Signal & Scenario & Events (raw) & Cut I & Cut II & Cut III &Cut IV  & ${\cal(S/\sqrt B)}
		_{100\,{\rm fb}^{-1} (1000\, {\rm fb}^{-1}) [3000\, {\rm fb}^{-1}]} $ \\
		\hline
		\hline
		$\nu_e H^{\pm} b$&I-110 & 2562 & 298 & 182 & 134 & 54  & 1.43 (4.52) [7.82]\\
				&I-130 & 1300 & 139 & 82 & 64 & 19  & 0.58 (1.82) [3.16]\\
				&I-150 & 347 & 29 & 13 & 11 & 3 & 0.16 (0.5) [0.86]\\
				&I-170 & 13 & 1.29 & 0.62 & 0.51 & 0.14  & 0.01 (0.03) [0.05]\\
                     \hline
		$\nu_e H^{\pm} b$&II-110 & 2183 & 245 & 151 & 122 & 53  & 1.4 (4.43) [7.68]\\
				&II-130 & 1128 & 128 & 84 & 71 & 22  & 0.7 (2.21) [3.82]\\
				&II-150 & 294 & 28 & 14 & 13 & 4  & 0.2 (0.65) [1.13]\\
				&II-170 & 6 & 0.6 & 0.33 & 0.3 & 0.08  & 0.005 (0.017) [0.029]\\
		\hline
		$\nu_e H^{\pm} b$&Y-110 & 6417 & 468 & 567 & 347 & 156 & 4.18 (12.99) [22.5]\\
				&Y-130 & 3268 & 366 & 204 & 156 & 46 & 1.43 (4.53) [7.84]\\
				&Y-150 & 847 & 68 & 29 & 23 & 6  & 0.33 (1.06) [1.83]\\
				&Y-170 & 22 & 2.3 & 1.12& 0.89 & 0.25  & 0.017 (0.05) [0.09]\\
		\hline
				$\nu_e bb j$& & 20169 & 2011 & 748 & 569 & 125&\\
		\cline{1-7} 
		$\nu_e b jj$& & 117560 & 10278 & 7211 & 5011 & 718 & ${\cal B}=1441$\\
		\cline{1-7}
		$\nu_e b t$&  & 41885 & 2278 & 1418 & 1130 & 188 & $\sqrt{{\cal B}}=37.9$\\
		\cline{1-7} 
		$\nu_e jjj$&  & 867000 & 9238 & 3221& 2593 &  409 &\\ 
		\hline
		\end{tabular}
		}
\caption{Significances obtained  after the sequential cuts described in the text for the signal process  $e^-q\to \nu_e H^- b$  followed by {$H^-\to b \bar{c}$} for four BPs in the  2HDM-III  like-I, -II and -Y. The simulation is done at detector level. {In the column Scenario,  the label A-110(130)[150]\{170\}  means  $m_{H^\pm} =110$(130)[150]\{170\} GeV in the 2HDM-III like-A, where A can be I, II and Y.} }
		\label{cb2}
\end{table}

\subsection{The  process  $e^-q\to \nu_e H^- b$  with $H^-\rightarrow \tau \bar \nu_\tau$ in the 2HDM-III like-X}

Now we focus our attention on the channel $H^-\to \tau\bar \nu_\tau$. To this effect, we look at leptonic $\tau$ decays ($\tau\to l\bar\nu_l \nu_\tau$, with $l=e,\mu$).

\begin{enumerate}[I]

\item This first set of cuts is focused on selecting events with one $b$-tagged jet and one lepton, by imposing $|\eta(b_{\rm tag},l)|<2.5$, $P_T(b_{\rm tag},l)>20$ GeV and the isolation condition $\Delta R (b_{\rm tag};l)>0.5$. 

\item The next set of cuts enables us to select a stiffer lepton and impose conditions on the missing transverse energy which are adapted to the trial $H^\pm$ mass. We select events with $P_T(l)>25(40)$ GeV and $E_T\hspace{-4mm}/>30(40)$ GeV for $m_{H^\pm}=110$, $130$($150, 170$) GeV.

\item We impose the cut $|\eta(b_{\rm tag})|>0.5$. Furthermore, upon defining the total hadronic transverse energy $H_T=\sum_{\rm hadronic}|P_T|$ in the final state, we select $H_T<60$ GeV.

\item Finally, we enforce the las selection by exploiting the transverse mass $M_T(l)^2=2 p_T(l) E_T\hspace{-4mm}/~~(1-\cos\phi)$, where $\phi$ is the relative azimuthal angle between $p_T(l)$ and  $ E_T\hspace{-4mm}/$~~, a quantity which allows one to label the candidate events reconstructing the charged Higgs boson mass. We make the following selection: $ m_{H^\pm} -50$ GeV $ <M_T(l)  <m_ {H^\pm} + 10 $ GeV.
\end{enumerate}

\begin{table} \label{taunu}
\centering
\resizebox{10cm}{!} {
		\begin{tabular}{|r|c|c|r|r|r|r|c|}
		\hline
		\hline
		Signal & Scenario & Events (raw) & Cut I & Cut II & Cut III &Cut IV & ${\cal(S/\sqrt B)}
		_{100\ {\rm fb}^{-1}(1000\ {\rm fb}^{-1})[3000\ {\rm fb}^{-1}]} $ \\
		\hline
		\hline
		$\nu_e H^- q$ &X-110 &6480 & 178 & 124 &94 &67 &2.41 (7.61) [13.19] \\ 
			&X-130 &3390 & 75 & 54& 52 & 35 & 1.13	 (3.58) [6.2]\\
			&X-150 &880 &6 & 3& 2 & 2& 0.09 (0.29) [0.5]\\
			&X-170 & 20 & 0.4 & 0.3 & 0.2 & 0.09 & 0.01 (0.02) [0.04]\\
		\hline
		$\nu_e b b j$ &      &20170 &  85& 56 & 23 &13 &  \\ 
		\cline{1-7}
		$\nu_e bjj $ &       & 117559&   623& 340 & 122 & 84 & \\ 
		\cline{1-7}
		$\nu_e tb $ &        &48845 &  460 & 374 & 149 & 105& ${\cal B}=763$  \\ 
		\cline{1-7}
		$\nu_e jjj$ &     & 867000 &  981 & 596 & 267 & 162 & $\sqrt{\cal B}=27.62$ \\ 
		\cline{1-7}
		$\nu_e l \nu_l j$ &    &23700   & 29 & 26&  8& 5&  \\ 
		\cline{1-7}
		$\nu_e l \nu_l b$ &    & 40400 & 1500 & 1203 & 569& 392&  \\ 
		\hline
		\end{tabular}
		}
		\caption{Significances obtained  after the sequential cuts described in the text for the signal process  $e^-q\to \nu_e H^- b$  followed by {$H^-\to \tau\bar \nu_\tau$}  for four BPs in the  2HDM-III  like-X. The simulation is done at detector level. {In the column Scenario,  the label X-110(130)[150]\{170\} means  $m_{H^\pm} =110$(130)[150]\{170\} GeV in the 2HDM-III like -X.} }
\end{table}

\section{Conclusions}
 
 Following the application of cuts I--IV, we obtain the signal and background rates in Tab. \ref{cb2}, for the 2HDM-III like-I, -II and -Y incarnations, and Tab. 4, for the like 
 X case. Statistically, significances of the signal ${\cal S}$ over the cumulative background ${\cal B}$ are very good at low $H^\pm$ masses already for 100 fb$^{-1}$ of
luminosity. Hence,  we confirm that the prospects for light $H^-$ detection in the 2HDM-III at the LHeC are excellent.

\end{document}